\documentstyle[epsfig]{elsart}

\newfont{\cmsy}{cmsy10}

\begin{document}

\begin{frontmatter}

\title{
Search For Disoriented Chiral Condensates In 158 AGeV Pb+Pb
Collisions
}

\author[inst13]{M.M.Aggarwal},
\author[inst14]{A.Agnihotri},
\author[inst20]{Z.Ahammed},
\author[inst]{A.L.S.Angelis}, 
\author[inst15]{V.Antonenko}, 
\author[inst6]{V.Arefiev}, 
\author[inst6]{V.Astakhov},
\author[inst6]{V.Avdeitchikov},
\author[inst12]{T.C.Awes},
\author[inst5]{P.V.K.S.Baba},
% ...

\collab{WA98 Collaboration}

\address[inst13]{University of Panjab, Chandigarh 160014, India}
\address[inst14]{University of Rajasthan, Jaipur 302004, Rajasthan,
  India}
\address[inst20]{Variable Energy Cyclotron Centre,  Calcutta 700 064,
  India}
\address[inst]{University of Geneva, CH-1211 Geneva 4,Switzerland}
\address[inst15]{RRC (Kurchatov), RU-123182 Moscow, Russia}
\address[inst6]{Joint Institute for Nuclear Research, RU-141980 Dubna,
  Russia}
\address[inst12]{Oak Ridge National Laboratory, Oak Ridge, Tennessee
  37831-6372, USA}
\address[inst5]{University of Jammu, Jammu 180001, India}
% ...

\newpage
\begin{abstract}
The restoration of chiral symmetry and its subsequent breaking through
a phase transition has been predicted to create regions of Disoriented
Chiral Condensates (DCC).
This phenomenon has been predicted to
cause anomalous fluctuations in the relative production of
charged and neutral pions in high-energy hadronic and nuclear collisions.
The WA98 experiment has been used to measure charged and photon
multiplicities in the central region of 158 AGeV Pb+Pb 
collisions at the CERN SPS.
In a sample of 212646 events, no clear DCC signal can be distinguished.
Using a simple DCC model, we have set a 90\% C.L. upper limit on the maximum 
DCC production allowed by the data.
\end{abstract}
\end{frontmatter}
\section{Introduction}
The approximate chiral symmetry of the QCD vacuum is believed
to be spontaneously broken in nature by the formation of an
isoscalar quark condensate.
Disoriented Chiral Condensates (DCC) may form in 
large, hot regions of hadronic matter where
this symmetry
has been briefly restored
\cite{raj95}. 
A DCC has an equal probability to be in any state 
related to the normal vacuum by
a chiral rotation.
By projecting the space of these available states onto a basis of
definite isospin,
it has been found that the charge distribution of 
pions emitted from a DCC has a characteristic
form \cite{anselm}:
\begin{equation}
\label{pf}
P(f)=\frac{1}{2\sqrt{f}}
\end{equation}
where $f$ is the neutral fraction,
\begin{equation}
\label{fdef}
f=\frac{ N_{\pi^{o}} }{ N_{\pi^{o}} + N_{\pi^{+}} +
  N_{\pi^{-}} }.
\end{equation}
This allows the possibility of hadronic interactions 
with anomalous fluctuations
between charged pions and neutral pions, as seen through their two-photon
decay channel.

The phenomenology of DCCs was first introduced in the context of
hadronic collisions by Bjorken et al, whose 
``Baked Alaska'' model \cite{alaska1} postulated that a hot shell,
expanding at the speed of light, could shield the cool interior from
the influence of the normal vacuum outside, allowing a large
region of DCC to form.
Rajagopal and Wilczek \cite{raj93a,raj93b}
studied the production of DCCs in nuclear collisions by
studying the chiral phase transition in QCD,
via its similarity to the O(4) Heisenberg magnet \cite{O4}.
Through numerical simulations,
they found that as the system rapidly expands and cools through 
the phase transition,
the equations of motion induce 
a non-equilibrium relaxation of the
chiral fields which amplifies
the production of soft pion
modes in a well-defined chiral direction.
This effectively creates clusters of low-$p_T$ pions,
with the cluster charge distribution following equation 
(\ref{pf}).
It should be noted that, in both studies, 
the strongest influence on the final state
composition is the symmetry itself rather than the exact
physics scenario studied.  Further work confirmed
these initial results, even after accounting for quantum fluctuations,
and proposed other mechanisms which might allow for large,
long-lived DCCs\cite{raj95}.

By allowing the possibility of events with almost no
electromagnetic energy, DCCs are an attractive hypothesis to
explain the ``Centauro'' events seen in cosmic rays
\cite{cosmic}.
These events have already motivated 
searches for unusual charge fluctuations
at the $S\overline{p}pS$ (by
UA1 \cite{ua1} and UA5 \cite{ua5}) and at the Tevatron (by Minimax
\cite{minimax} and CDF \cite{cdf}).
And yet, there have been no systematic studies utilizing
the simultaneous measurement of charged and neutral multiplicities
in heavy ion collisions at any energy.
It has been argued \cite{raj95} that 
heavy ion collisions at SPS energies, the highest currently available,
might create the large volumes
which favor the development of long-wavelength oscillations within the
reaction zone.
It is true that large baryon number in the central region 
complicates theoretical calculations and 
may obscure the initial signal via the rescattering of secondaries.
It is also possible that the low-energy observation that most pions
are produced resonantly via the $\rho$ and $\omega$ channels
leaves few ``direct'' pions which may be influenced by a DCC.
We must keep in mind, however, that these are extrapolations from
lower energies.  
Their cumulative effect is uncertain, especially at higher energies
where the formation of a quark-gluon plasma 
would render previous measurements inapplicable.
In any case, in the absence of any substantial experimental
evidence for or against DCCs in heavy ion collisions at SPS energies,
and a great deal of theoretical evidence in their favor,
it is imperative to simultaneously measure charged and neutral particles
at the SPS
and analyze their fluctuations to perhaps isolate a DCC signal.  Observing
such a signal might be an indication of the chiral phase transition
in hot nuclear matter.
\section{Experimental setup}
\begin{figure}
\begin{center}
\setlength{\unitlength}{1cm}
\begin{picture}(14,7)
\put(0,0){\framebox(14,7){
\epsfig{figure=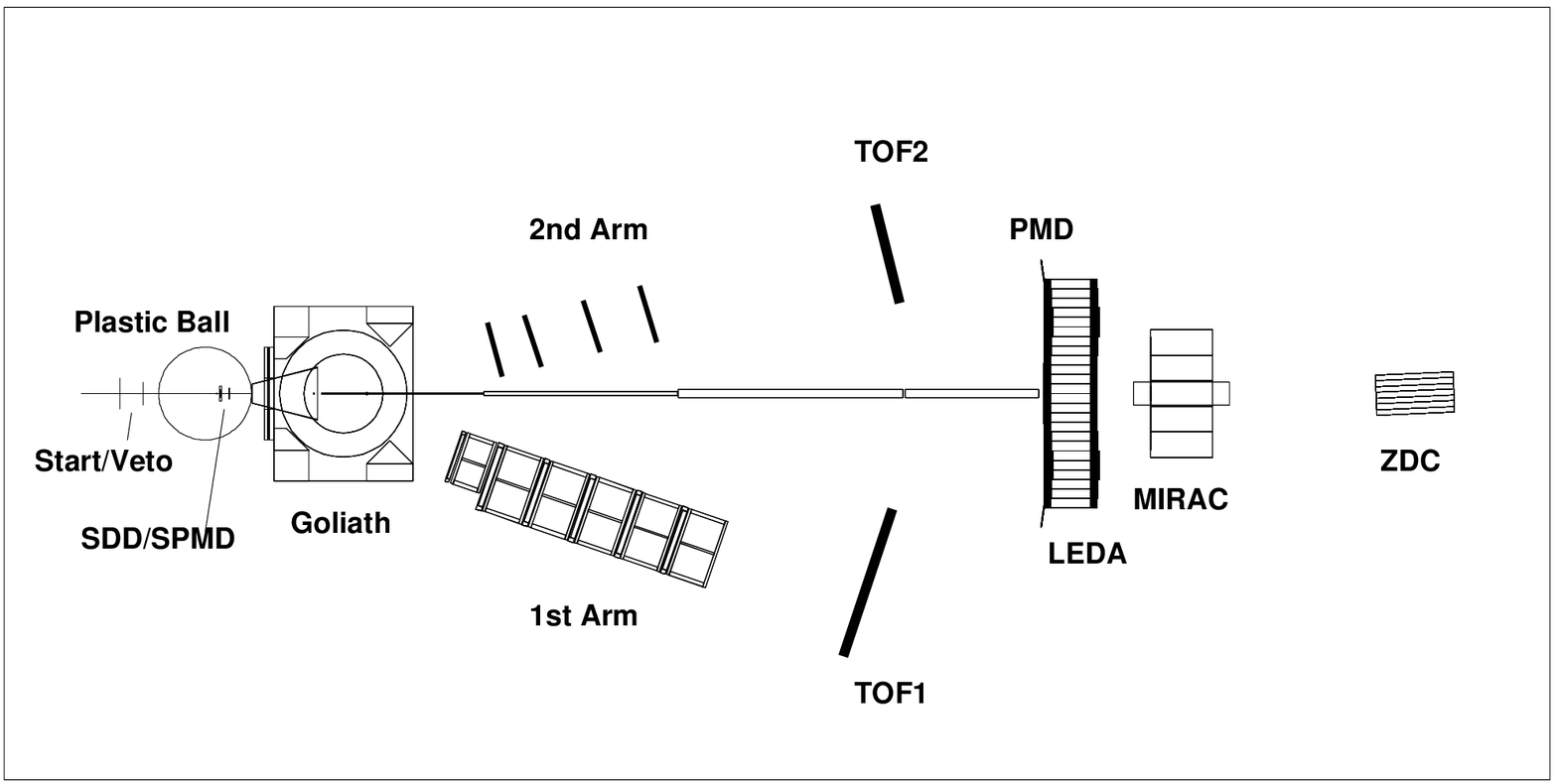,width=14cm,clip=true}}}
\end{picture}
\caption {\label{wa98sch} 
The WA98 Experiment at the CERN SPS.  This analysis uses the Silicon Pad
Multiplicity Detector (SPMD) and the Photon Multiplicity Detector (PMD)
to measure the charged and neutral multiplicity for each event,
and the Mid-Rapidity Calorimeter (MIRAC) 
for the measurement of event centrality.
}
\end{center}
\end{figure}

The WA98 experiment \cite{wa98} is a general-purpose,
large-acceptance photon and hadron spectrometer with the ability to
measure several different global observables event-by-event.  
For this search, we use a subset of the full apparatus,
shown schematically in Figure \ref{wa98sch}.
We measure charged particles with a Silicon Pad Multiplicity Detector
(SPMD) and photons with a Photon Multiplicity Detector (PMD).
Using these, we are able to count charged particles and photons 
in the central pseudorapidity region on an event-by-event basis.  
For a determination of the centrality of each collision, we use
the transverse energy ($E_T$) measured in the Midrapidity Calorimeter
(MIRAC \cite{mirac}).  For removal of background events, we also use the 
Zero-Degree Calorimeter (ZDC) and the Plastic Ball detector \cite{pball}.
\subsection{Charged particle multiplicity}
We count charged particles using a circular 
Silicon Pad Multiplicity 
Detector (SPMD) 
\cite{sp} located 32.8 cm from the target 
covering $2.35< \eta < 3.75$,
the central rapidity region at SPS energies (where $\eta_{CMS} = 2.9$),
and full azimuth.
The detector consists of four overlapping quadrants, each fabricated
from a single 300~{$\mu m$} thick silicon wafer.
The active area of each quadrant 
is divided into 1012 pads forming 46 azimuthal
wedges and 22 radial bins with a pad size increasing with 
radius to provide uniform pseudorapidity coverage.
The efficiency of detecting a charged
particle in the active area has been determined in a test beam to be
better than 99{\%}.  Conversely, the detector is transparent to 
high energy photons, since only about 0.2{\%} are expected to
interact in the silicon.
During the data taking, 95\% of the pads worked properly and are
used in this analysis.

In a central ion-ion collision, the occupancy can be as high as
$20\%$,
implying that $\approx20\%$ of the pads contain two or more hits.
An unbiased way to estimate the total number of charged particles
in a given event
under such conditions is to use the sum of the energy deposited
in pads exceeding 1/2 of the most probable energy loss
divided by the mean energy loss per
particle as determined
in low-multiplicity events: 
\[
N_{ch} = \sum_{i={\mathrm{hits}}}\frac{{dE/dx}_{i}}
{\langle dE/dx \rangle}.
\]
Because of the fluctuations in the energy loss, 
described by a Landau distribution,
the uncertainty on $N_{ch}$ can be estimated to
be $\Delta N / N = {60\%}/\sqrt{N}$.
For typical central events with a multiplicity of $\approx 600$,
this gives an uncertainty of about 2\%.
To check the overall scale, we
compare the results with the multiplicity obtained by assuming
that the particles are distributed uniformly so the multi-hit
probability is given by Poisson statistics.  A simple calculation
gives $N_{\mathrm{ch}}' = -N_{\mathrm{pads}} 
\log (1-N_{\mathrm{hits}}/N_{\mathrm{pads}})$,
where $N_{\mathrm{pads}}$ is the total number of pads, 
and $N_{\mathrm{hits}}$ is the total number of hit pads.
Using this as a check, we estimate the systematic 
error on $N_{ch}$, due to uncertainties in the gains
and backgrounds, to be about 4\%.  

\subsection{Photon multiplicity}\label{subsec:phot}
We count photons in the preshower Photon Multiplicity Detector (PMD)
situated 21.5 m from the target, covering the region
$2.8 < \eta < 4.4$. The photons
impinging on the detector are converted in  3.34 $X_0$ thick lead
and iron and the secondaries are detected in
3mm-thick square plastic scintillator pads of varying sizes
(15mm, 20mm and 23mm). 
A matrix of
50 $\times$ 38 pads is placed in one light-tight box module and read 
out individually via wavelength shifting optical fibers coupled to
an image intensifier and CCD camera system 
similar to that described in \cite{pmd}.
The modules with smaller pads were mounted in the forward angle region to 
minimize cluster overlap at large multiplicities and to provide
reasonably uniform occupancy. 
Out of a total of 28 box modules implemented in
the PMD, the data presented here correspond to
19 box modules having 35524 pads. 
The average occupancy 
for the part of the detector considered in the present case is around 15\% for 
central events.

The principle of photon identification makes use of the fact that photons are 
more likely  to shower in the lead converter and produce a large signal in 
the scintillator pads, while non-showering hadrons will produce a signal
corresponding to a single minimum ionizing particle (MIP). Signals from
several neighbouring pads are combined to form clusters and those with energy
deposition larger than that corresponding to 3 MIPs are considered to be
"$\gamma$-like" clusters. This selection gives an average photon 
counting efficiency  of about 70\% which is almost uniform over the 
range of centrality and pseudorapidity considered.
It also creates an effective lower $p_T$ cutoff of 30 MeV/c, at which
point the efficiency falls below 35\%.
About 15\% of the produced hadrons
impinging on the PMD interact in the converter, generating secondaries
which also deposit large energy on the detector. 
This 
contamination constitutes a background to photon counting. In
order to minimize effects due to variations in the angular distributions
of charged particles,
we only use data with the Goliath magnet turned off.

The photon counting efficiency, hadron contamination and the associated errors
are derived using test beam data and GEANT simulation using a method similar
to the ones described in \cite{pmd,pmdmult}. 
The level of hadron contamination in the PMD was verified
by comparing the azimuthal distribution of hits for magnet-on
and magnet-off data\cite{erik}. 
The azimuthal distribution of charged tracks becomes very non-uniform
in the presence of the magnetic field, the amount of non-uniformity  
indicating the magnitude of the hadron contamination. 

It should be emphasized
that in this analysis, we do not correct the data using these
parameters. Instead we account for all of the detector effects by fully 
simulating the conversion of particles in the detector, as described below.
\subsection{Data and Event Selection}\label{subsec:evsel}
In this analysis, we study reactions induced by a 158 AGeV
Pb beam incident upon a 213$\mu$m thick \nuc{208}{Pb} target.
The fundamental ``beam'' trigger condition consists of a signal in
a gas \v{C}erenkov start counter \cite{start} located 3.5 meters upstream
of the target and no coincident signal in a
veto counter with a 3mm circular hole located 2.7
meters upstream from the target.
A beam trigger is considered to be a minimum-bias interaction if
the transverse energy sum in the full MIRAC acceptance
exceeds a lower threshold.

Pileup events are eliminated using a system of TDCs 
each started by a particular trigger counter and stopped by a 
second trigger.  
Using these, we remove events where a second interaction occurred
within a $\pm 10\mu s$ window before and after the recorded event.
Still, our TDC system cannot distinguish two events that 
arrive less than 50 ns apart.
These are eliminated by requiring the
sum of energy deposited in the MIRAC ($3.5<\eta<5.5$) and 
ZDC ($\eta>6$) to be
consistent with a single event.
After applying these cuts, 70\% of the data sample remains.
\section{General Features of Data and Comparison with VENUS 4.12}
To describe the bulk of the data, we use the VENUS 4.12 \cite{venus}
event generator with its default settings.
To compare VENUS with our data, we propagate
the raw generator output through a 
full simulation of our experimental setup using the 
GEANT 3.21 \cite{geant} package from CERN.
The simulation incorporates the detector physics
effects
and folds them into the generated data, which is then
analyzed using the same code used for the raw experimental data.
In the rest of this paper, the term ``VENUS'' refers 
to the combination of VENUS 4.12 and
the full GEANT 3.21 detector simulation, not to the raw generator
output, unless otherwise specified.

The SPMD simulation includes the effect of Landau fluctuations in the
energy loss of charged particles in the silicon and the pad geometry
of the detector.
In addition to the secondaries from the ion-ion collision itself,
the SPMD is also sensitive to the
$\delta$-rays generated by the \nuc{82+}{Pb}
ion passing through the lead target.  We can get a conservative 
estimate of the $\delta$-ray multiplicity in physics events by studying
events that satisfy the conditions for a beam trigger but not the
interaction trigger.  These ``beam'' events have a mean multiplicity
in the SPMD of 11.4$\pm$.5 and a width of 5.9$\pm$.3.  
The angular distribution is consistent
with a spatially uniform illumination of the detector surface.
To include these ion-induced $\delta$-rays in the simulation, 
we sample the measured
charged multiplicity distribution for beam
events and add it to the charged particle multiplicity
for each simulated event.
We estimate the uncertainty in the absolute scale of $N_{ch}$ from
the simulation to be less than 3\% and the relative uncertainty 
between data and VENUS to be less than 2\%.

The PMD simulation also incorporates the effects of additional fluctuations 
to the energy loss arising due to the statistical nature of the 
scintillation process; light transport through the wavelength shifting fibres
and the image intensifier chains; and imperfections in the electro-optical 
imaging. 
The widths due to this extra fluctuation were obtained by a comparison of the
GEANT and test beam results using single pions and electrons at various 
energies. As all of the readout 
chains were not used in the test beam experiment,
a method of detailed intercomparison of the various features of data and
simulation was used to obtain
the gains of the
individual readout units.
We estimate the uncertainty on the absolute multiplicity scale of 
simulated $\gamma$-like clusters, due to uncertainties
in various parameters of the simulation and data analysis, to be 15\%, 
and that the relative uncertainty between data and VENUS is 5\%.

\begin{figure}
\begin{center}
a.)\epsfig{figure=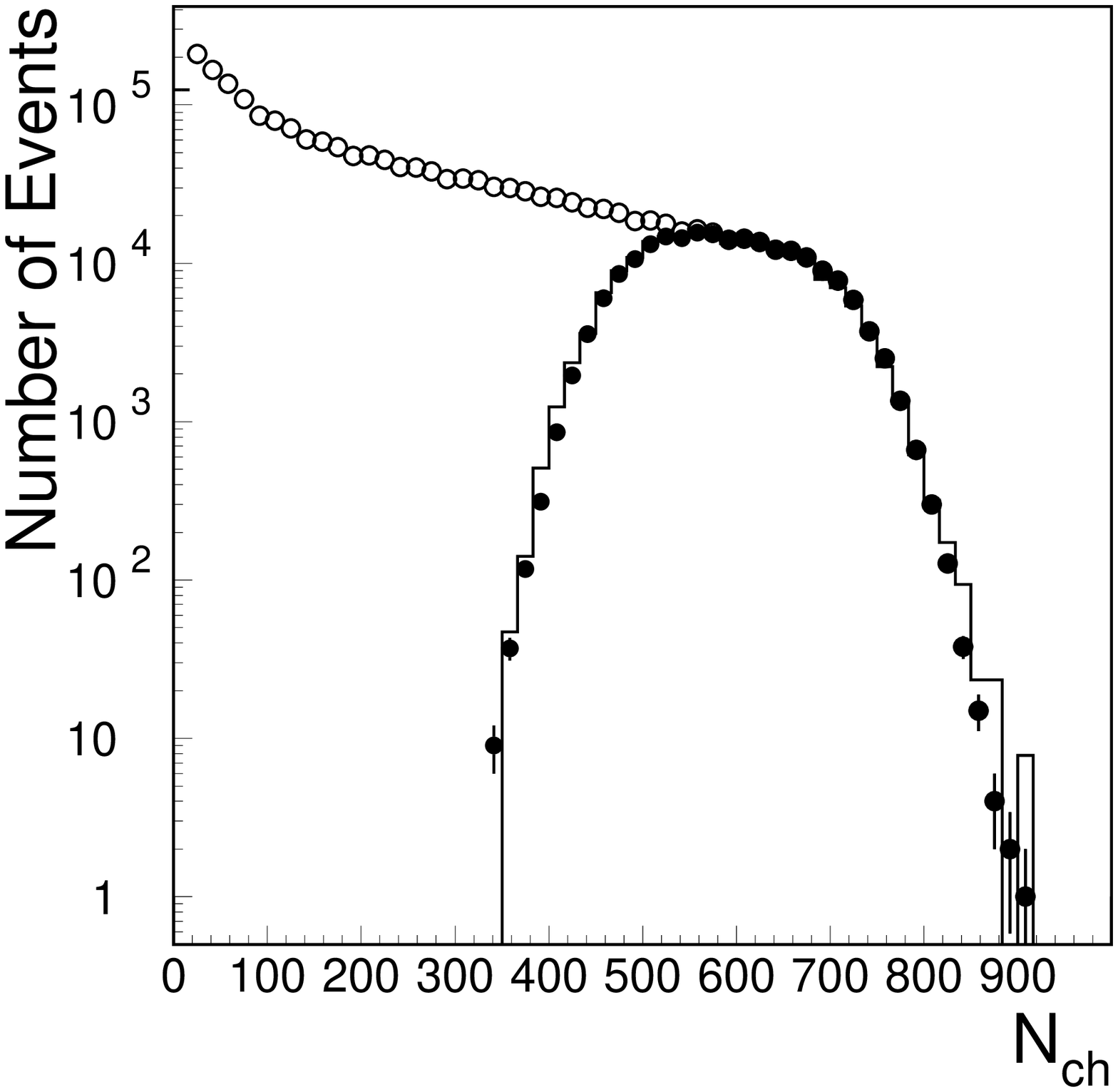,width=10cm}
b.)\epsfig{figure=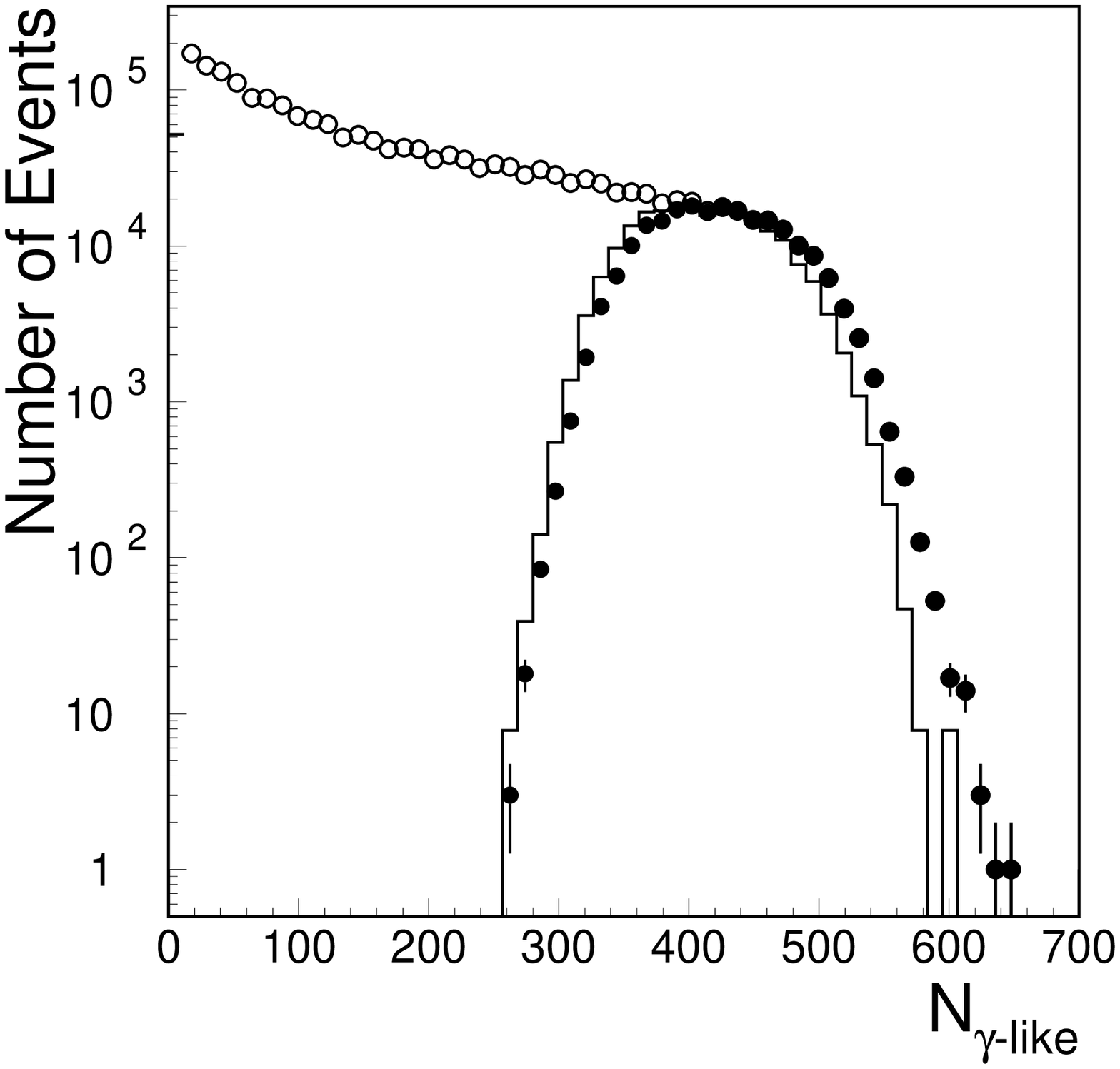,width=10cm}
\caption {\label{mult} 
The charged and neutral multiplicity distributions are shown in a) and
b).  
The open circles represent the minimum-bias distribution.  
The ``central'' sample ($E_T > 300$ GeV) is represented by
closed circles for the data and by histograms for VENUS.
}
\end{center}
\end{figure}

\begin{figure}
\begin{center}
\epsfig{figure=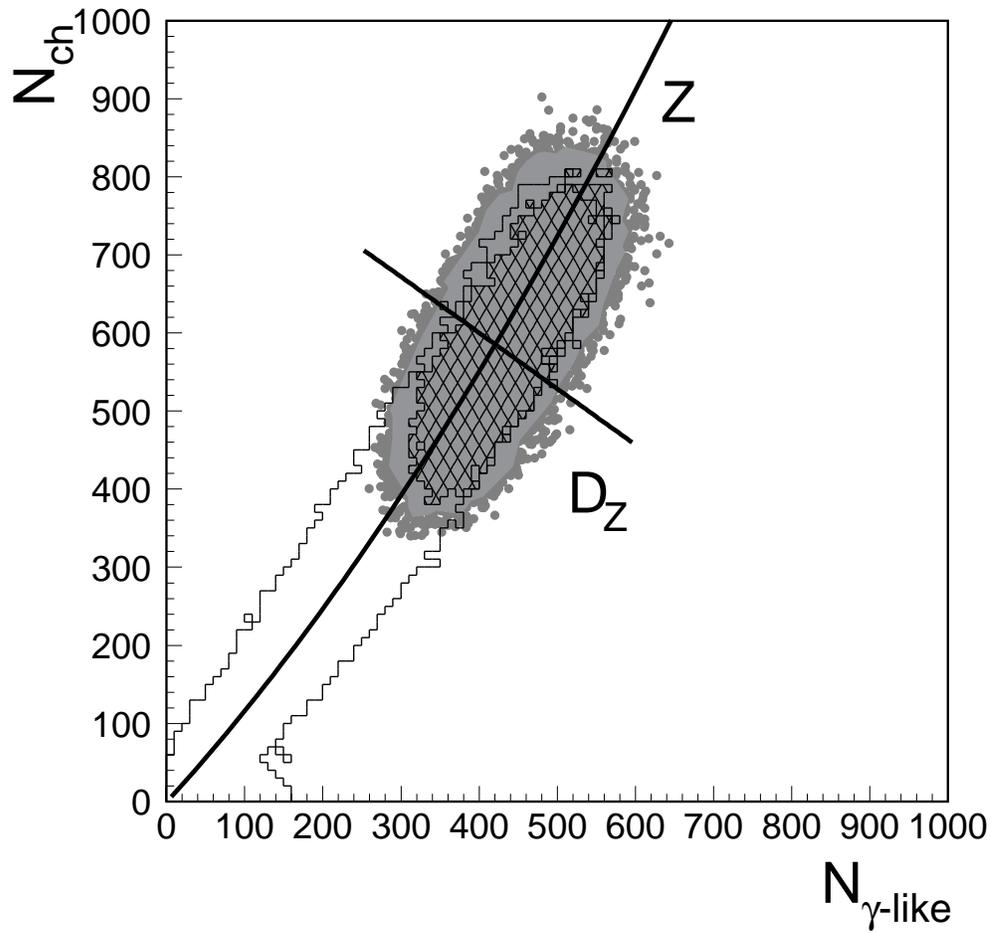,width=14cm}
\caption {\label{96data} 
This is the scatter plot showing the correlation between $N_{ch}$
and $N_{\gamma-{\mathrm{like}}}$.  
The solid outline shows the trend of the minimum
bias data.  The central sample (with $E_T>300$ GeV) is shown
as points for the data and as
a hatched region for VENUS (with much lower statistics).
Overlaid on the plot are the Z axis and the $D_Z$ axis at a particular
value of Z as explained in the text.
}
\end{center}
\end{figure}

In Figures \ref{mult}a and 
\ref{mult}b we present the minimum-bias multiplicity
distribution for charged particles and $\gamma$-like clusters. 
For the DCC search,
we will concentrate on the 10\% most central events, defined by a measured
transverse energy of at least 300 GeV in $3.5<\eta<5.5$.
These correspond roughly 
to the top 620 mb of the Pb+Pb minimum bias cross 
section $\sigma_{mb}=$6200 mb.
After all cuts are applied, there are 212646 events in this sample,
which we will refer to as the ``central'' sample in the rest of this
paper.
The central data sample is shown by closed circles in Figs. \ref{mult}a
and \ref{mult}b and a comparison with VENUS events chosen by identical
cuts is shown by the histogram.
The correlation between the charged and neutral multiplicities is
presented in Figure \ref{96data} with the minimum bias distribution
outlined, the central VENUS events hatched, and the central data events 
shown as scattered points, each point corresponding to a single event.

The most distinctive feature of the scatter plot is the strong correlation 
between the 
charged and neutral multiplicities.
A reasonable explanation of this would be if most of the produced
particles are pions with their charge states partitioned binomially,
as measured in $pp$ experiments at similar energies \cite{pp1}.
A binomial distribution leads to a correlation
width $\sigma(N_{ch}-N_{\gamma})
\propto \sqrt{N_{ch}+N_{\gamma}}$, which would explain the very
tight correlation, since the relative fluctuations are proportional to
$1/\sqrt{N_{ch}+N_{\gamma}}$.
As this is seen in both data and VENUS, we can study the
contributions to the different multiplicities to verify 
this hypothesis.  In fact, 
about 80{\%} of the charged particles produced in VENUS are pions, the
rest being protons and kaons.  Moreover, about 85{\%} of produced 
photons come from $\pi^{o}$ decays.
Thus, by simply counting the charged particles and
photons produced in a heavy ion collision, we have a reasonable
estimate of the number of charged and neutral pions created.

We verify the binomial 
nature of the charge fluctuations in VENUS by
studying its ``binomiality'':
\begin{equation}
B = \frac {N_{\pi_{ch}}-p_{ch}N_{\pi}}{\sqrt{p_{ch}(1-p_{ch})N_{\pi}}}
\end{equation}
where $N_{\pi_{ch}}$ and $N_{\pi}$ are number of charged pions and
the total number of pions for each event, and $p_{ch}=N_{\pi_{ch}}/N_{\pi}$ is the probability
that a pion is charged.
For a pure binomial distribution, 
$p_{ch} = 2/3$ and B is Gaussian with a mean at zero
and an RMS of one.  
For VENUS without GEANT, we find an RMS of approximately $.95$ for pions 
produced in the central rapidity region in events with 
an impact parameter less than 6 fm.
This is consistent with
the hypothesis that the correlation arises mainly from the 
binomial partition of $N_{\pi}$, the total pion multiplicity.
\section{Event-by-Event Search for DCCs}
DCCs should modify the binomial partitioning
of $N_{\pi}$ into charged and neutral pions.
Events in which a DCC is produced 
(henceforth referred to as ``DCC events'') 
will show up as
deviations from the binomial behavior and appear
as outliers with respect to the bulk of the data.
We have already discussed that the charged and neutral multiplicities
are directly sensitive to the charged and neutral pion multiplicities
in each event.
Thus, DCC events should appear in the correlation of charged and
neutral multiplicities, while the individual distributions will
be mainly unaffected. 
\subsection{Data Analysis}
\begin{figure}
\begin{center}
a.)\epsfig{figure=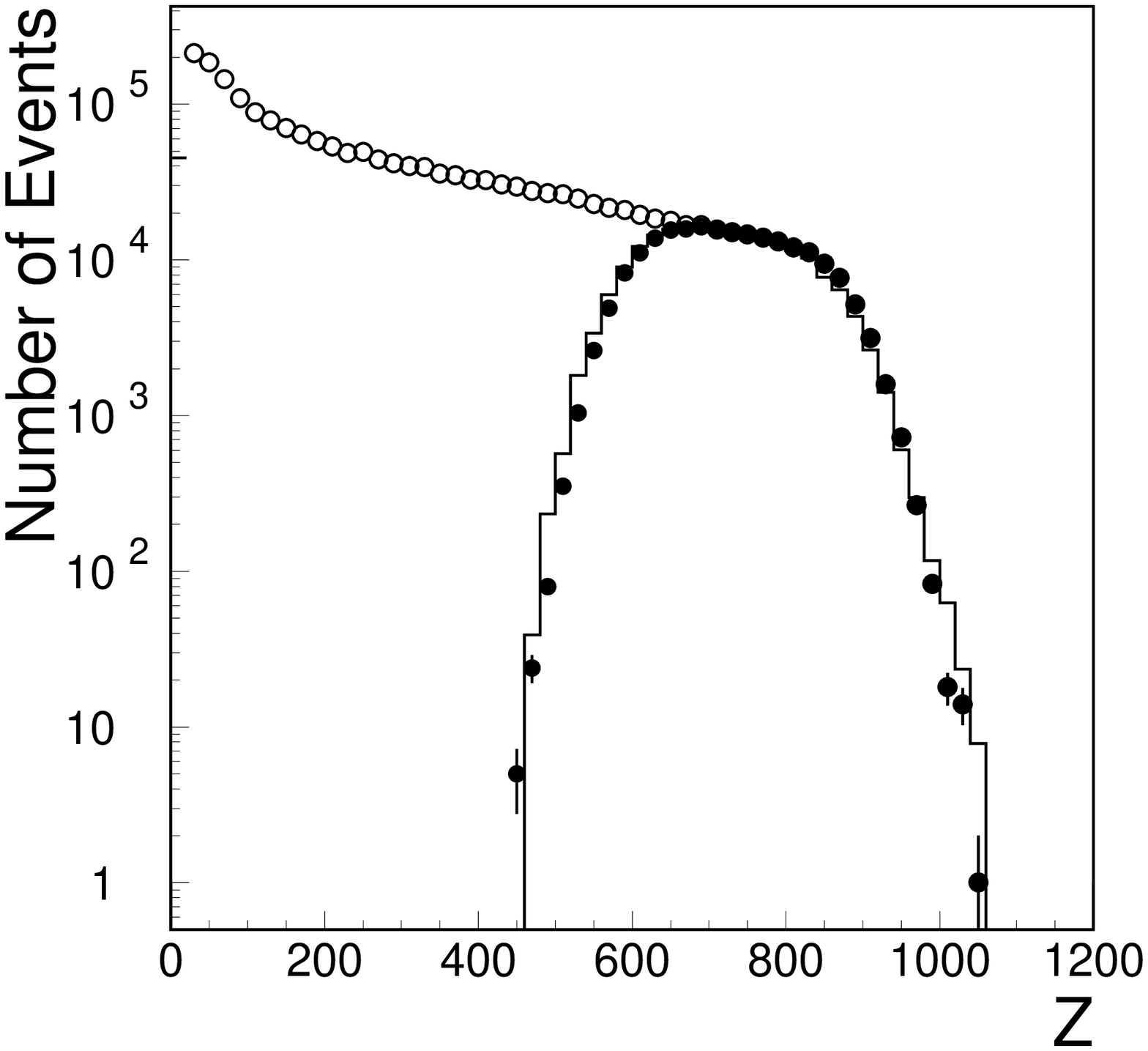,width=10cm}
b.)\epsfig{figure=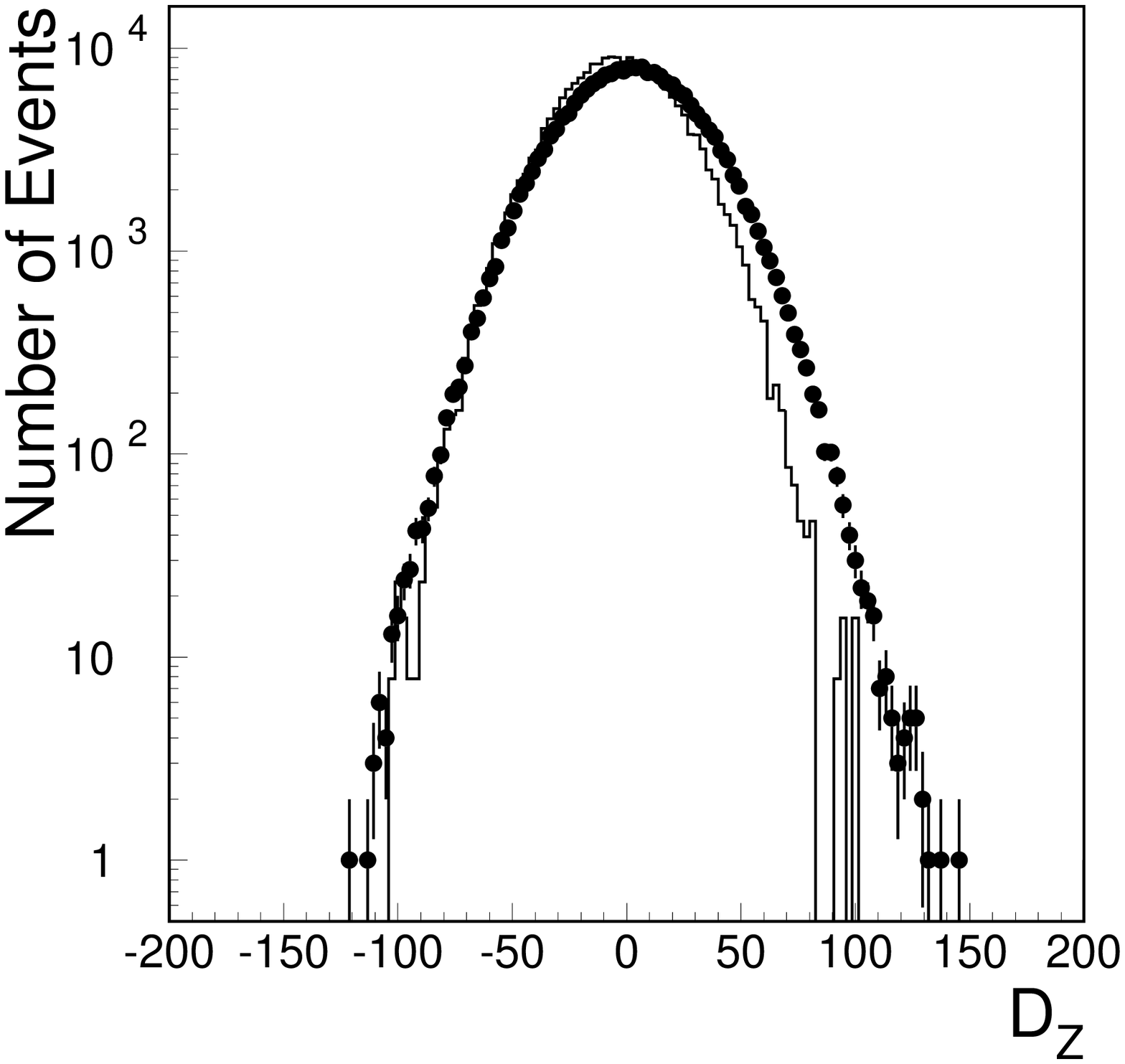,width=10cm}
\caption {\label{zproj}
a.) This figure shows the distribution of Z, with the same conventions as 
in figure \ref{mult}.  
b.) This shows the distribution of $D_Z$ in the ``central'' sample
for the data (closed circles) and for VENUS (histogram).  
The difference in the mean between the two distributions arises due to the
overall scale differences discussed in Section 3.
}
\end{center}
\end{figure}

The strong correlation between charged and neutral multiplicities 
described above suggests a more appropriate
coordinate system with one axis being the measured correlation axis
and the other perpendicular to it.
If all detected particles were pions and the detectors 
were perfect and had identical
pseudorapidity acceptance, then the correlation axis would
be a straight line.
Instead, we must account for the fact that at high multiplicities,
the pseudorapidity distributions tend to narrow, changing
the relative acceptance of charged and neutral particles 
due to the non-identical apertures of the SPMD and PMD.
Moreover, the large occupancies in the PMD lead to a slight
saturation effect.
It is then useful to define a coordinate system consisting 
of a correlation axis ($Z$) described by a second-order polynomial,
and the perpendicular distance ($D_Z$) from it, which is
defined to be positive for points below this Z axis.
These axes are shown superimposed on Figure \ref{96data}
and the projection along the Z-axis is shown in Figure \ref{zproj}a.
The full projection along the $D_Z$-axis is shown in Figure \ref{zproj}b.
To a very good approximation, the data are Gaussian distributed,
which is consistent with binomial partition.
The VENUS results, shown by the histogram, are also Gaussian,
but with a slightly smaller width.

In both cases, $\sigma_{D_Z}$, the 
standard deviation of a gaussian fit in the $D_Z$ direction, increases with
increasing Z.  We have chosen to work with the scaled variable
$S_Z{\equiv}D_Z/\sigma_{D_Z}$ in order to compare relative fluctuations
at different multiplicities.
While binomial partition leads to fluctuations that grow as
$\sqrt{N}$, the data and VENUS follow a slightly different power
law, due to the presence of contaminating particles, like nucleons
kaons, and etas.
A reasonable parametrization of $\sigma_{D_Z}$ for $Z>200$ has been found to be
$\sigma_{D_Z} = C + Z^{\beta}$
where $C=7.5\pm.1$ for the data, and $C=4.8\pm.1$ for the simulated events,
and $\beta$=.46.  The discrepancy between VENUS and the data 
can be seen more clearly by measuring the width of the $S_Z$ distribution
with the $\sigma_{D_Z}$ in the denominator taken from the simulation.  
The VENUS distribution is a gaussian
of width .998$\pm$.002 (fit error only) and the data is also gaussian,
of width 1.13$\pm$.07
(error from relative scale uncertainties included).
\subsection{Model of DCC production}
\begin{figure}
\begin{center}
\epsfig{figure=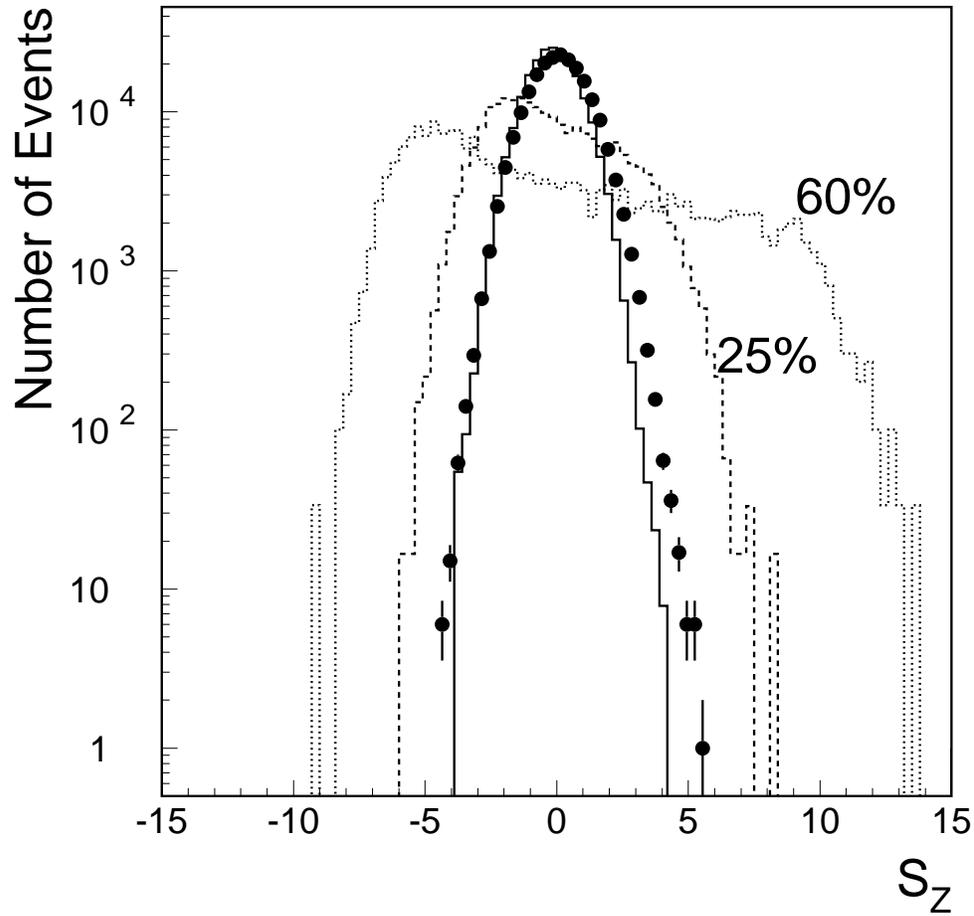,width=14cm}
\caption {\label{dcc_hyp}
$S_Z$ distribution for the experimental data is shown, 
overlaid with VENUS simulations incorporating 
0\%, 25\% and 60\% DCC in every event.
All of the distributions are normalized to
the total number of data events.  
}
\end{center}
\end{figure}

To estimate the effect of DCC production we have modified the
VENUS events
to include characteristic
fluctuations in the relative production of charged and
neutral pions.  
We assume that only a single domain of DCC is
formed in each central collision.  
A certain fraction $\zeta=N_{DCC}/N_{\pi}$ of the 
VENUS pions is associated with this domain
and a value of $f$ is chosen
randomly according to the distribution shown in equation \ref{pf}.
Then the charges of the pions are interchanged pairwise
($\pi^{+}\pi^{-}$ or $\pi^{o}\pi^{o}$)
until the charge distribution
matches the chosen value of $f$.
This simulates a DCC accompanied by the normal hadronic background
in a way that conserves energy, momentum, and charge.
The $S_Z$
distribution for the $\zeta$ = 0\%, 25\% and 60\% DCC hypotheses are
shown in Figure \ref{dcc_hyp}
with the data overlaid as closed circles.  
It is clear that the distributions get
wider as $\zeta$ is increased. 
\subsection{Upper Limit Calculation}
As seen in the previous section,
DCC events
will show up as non-statistical
tails on the $D_Z$ axis.
We see no such events in our data sample.  Thus, we are faced with
the possibilities that single-domain DCCs are very rare, very small, or both.
To check which hypotheses are consistent with our data,  we determine 
upper limits on the frequency of DCC production as a function of
its size, as represented by $\zeta$.  

We have computed $S_Z$ distributions
for several values of $\zeta$, ranging from 15\% to 90\%.
To define an efficiency for detecting DCCs, 
we start from the observation that the distribution 
assuming the null hypothesis is Gaussian.
With our statistics,
we expect few events farther than 5 to 6 $\sigma$ from the mean.
An event containing a DCC, however, has an enhanced probability of 
being found in this region.
The cut $|S_Z| > S_{\mathrm{cut}}$ then defines a
two-dimensional region in the scatter plot in which all events
are considered to be ``DCC candidates''.  
Once the cut is set,
the DCC efficiency is defined, for $N_{\mathrm{MC}}$ VENUS events, as
\begin{equation}
\epsilon(S_{\mathrm{cut}},\zeta) = 
\frac{N(|S_Z|>S_{\mathrm{cut}},\zeta)}{N_{\mathrm{MC}}}
\end{equation}
which is a function of both the DCC fraction and the cut position.

The background is determined by a Gaussian fit to the VENUS
distribution, in order to extrapolate beyond the Monte Carlo statistics.
With the efficiency and background determined,
we calculate the Poisson upper limit $N_{U.L.}$ for a 90\% confidence level,
which is $\approx$2.3 if there are no measured events 
over the cut and no  background events are expected.
These three numbers are combined into an upper limit, for 
$N_{\mathrm{Data}}$ events, via the formula:
\begin{equation}
\frac{N_{\mathrm{DCC}}}{N_{\mathrm{Central}}}(S_{\mathrm{cut}},\zeta)
\leq
\frac{N_{U.L.}}{\epsilon(S_{\mathrm{cut}},\zeta)}
\frac{1}{N_{\mathrm{Data}}}.
\end{equation}

\begin{figure}
\begin{center}
\epsfig{figure=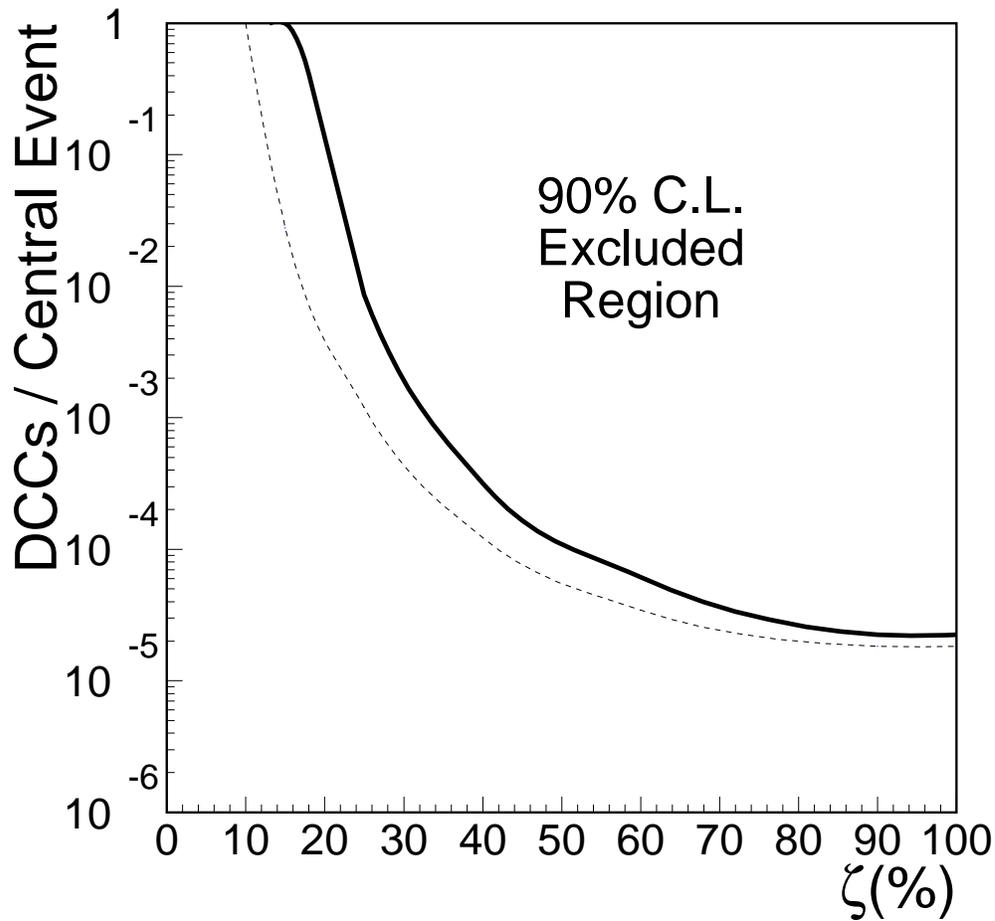,width=14cm}
\caption {\label{limits} 90\% C.L
upper limit on DCC production per central event as a function of
the fraction of DCC pions under two assumptions.  The thick line
gives the upper limit obtained by assuming the $\sigma_{D_Z}$ in 
$S_Z$ is completely given by the VENUS calculation requiring to
make a cut at 6$\sigma$. 
The dashed
line shows a less conservative limit obtained by 
using the $\sigma_{D_Z}$ measured in
the data itself.  This allows us to make a tighter cut
at 5$\sigma$, increasing the DCC detection efficiency.
}
\end{center}
\end{figure}

We have calculated limits for two scenarios.  The first is based
upon the conservative assumption that VENUS should
describe the data perfectly in the absence of a DCC signal.
Under these assumptions, $S_Z = D_Z/\sigma_{D_Z}$ as obtained
from VENUS
(as it was in Figure \ref{dcc_hyp})
and $S_{\mathrm{cut}}=6.$, which is well away from the
data point with the largest $S_Z$.  The 90\% C.L. limit is presented in 
Figure \ref{limits} as a solid line.  The other scenario assumes
that the difference between the data and VENUS is due to detector
effects and that the widths should be the same.  In this case,
$S_Z = D_Z/\sigma_{D_Z}$, with $\sigma_{D_Z}$ taken
from the data, and 
we choose a tighter cut $S_{\mathrm{cut}}=5$.  This
limit is presented in the same figure as a dashed line.  The two
limits are quite different at $\zeta = $15\% but
get closer at $\zeta >$ 30\%.
In both cases, the uncertainty in the absolute comparisons between
the data and VENUS have not been included in the upper limit estimate.
\section{Discussion}
Earlier studies estimated the DCC radius to be around $R\approx 3-4$ fm. 
Coupled with a vacuum energy density $u$ 
given by the chiral effective potential to
be 60-120 MeV/fm$^3$, and an assumed Gaussian $p_T$ 
distribution of width $\approx 1/R$,
an average DCC
was thought to generate $\frac{4}{3}\pi R^3 (u/m_\pi) \approx 50-230$
pions \cite{gavinmodel}.
The SPMD sees all of the
charged particles produced in the central rapidity region, 
80\% of which are pions, letting us
estimate the total number of centrally produced pions in 
an average event to be about 720.  Thus, we would expect a
DCC to be approximately $\zeta{\approx}5-30\%$.  Our analysis clearly
rules out anything larger than about 25\% within the
scope of the assumed model, but cannot say much about
anything smaller.  
However, if the pions tend to cluster in phase space, there
are methods that should be able to find DCC events, and these
are currently under study \cite{huang,tapan}.

A small and frequent DCC might also appear as a
slightly enhanced width,
similar to what we observe in our data when compared to VENUS.
However, this enhancement could
also result from uncertainties in the detector modelling or
the underlying physics model itself.
For instance, theoretical uncertainties might arise because no model
has ever been used to study charge correlations in heavy ion collisions.
Rescattering phenomena, resonances, or Bose-Einstein effects
may have
predictable effects on the expected binomial distribution. 
These issues will be addressed in future studies.
\section{Conclusions}
We have used the WA98 apparatus to perform the first 
search for the production of
Disoriented Chiral Condensates in 158 AGeV Pb+Pb
collisions.
No events with large charged-neutral fluctuations have been observed.
By comparing the correlations of the charged and neutral multiplicity,
measured on an event-by-event basis, to a simple model
incorporating a DCC signal into
VENUS 4.12 events, we have set a 90{\%} CL upper limit on the
frequency of DCC production as a function of its size.
\section{Acknowledgements}
We wish to express our gratitude to the CERN accelerator division for
excellent performance of the SPS accelerator complex. We acknowledge with
appreciation the effort of all engineers, technicians and support staff who
have participated in the construction of the this experiment. 
This work was supported jointly by 
the German BMBF and DFG, 
the U.S. DOE,
the Swedish NFR, 
the Dutch Stichting FOM, 
the Stiftung fuer Deutsch-Polnische Zusammenarbeit,
the Grant Agency of the Czech Republic under contract No. 202/95/0217,
the Department of Atomic Energy,
the Department of Science and Technology,
the Council of Scientific and Industrial Research and 
the University Grants 
Commission of the Government of India, 
the Indo-FRG Exchange Programme,
the PPE division of CERN, 
the Swiss National Fund, 
the International Science Foundation under Contract N8Y000, 
the INTAS under Contract INTAS-93-2773, 
ORISE, 
Research-in-Aid for Scientific Research
(Specially Promoted Research \& International Scientific Research)
of the Ministry of Education, Science and Culture, 
the University of Tsukuba Special Research Projects, and
the JSPS Research Fellowships for Young Scientists.
ORNL is managed by Lockheed Martin Energy Research Corporation under
contract DE-AC05-96OR22464 with the U.S. Department of Energy.
The MIT group has been supported by the US Dept. of Energy under the
cooperative agreement DE-FC02-94ER40818.
In addition we would like to thank
R. Birgeneau, H.Y. Chang, A.E. Chen, 
W.S. Edgerly, W.T. Lin, O. Runolfsson and B. Wadsworth.
\bibliography{WA98-3_final}
\newpage
\newpage
\newpage
\newpage
\newpage
\newpage
\newpage
\end{document}